\documentclass[letterpaper, aps, showpacs,onecolumn]{revtex4}

\usepackage{amsmath}
\usepackage{amssymb}
\usepackage{latexsym}
\usepackage{graphicx,epstopdf}
\usepackage{hyperref}

\begin{document}

\title{An extremal fractional Gaussian\\
with a possible application to option-pricing\\with skew and smile}

\author{Alexander Jurisch}

\affiliation{ajurisch@ymail.com, Munich, Germany}

\begin{abstract}
We derive an extremal fractional Gaussian by employing the L\'evy-Khintchine theorem and L\'evian noise. With the fractional Gaussian we then generalize the Black-Scholes-Merton option-pricing formula. We obtain an easily applicable and exponentially convergent option-pricing formula for fractional markets. We also carry out an analysis of the structure of the implied volatility in this system.

\end{abstract}

\maketitle

 
\section{Introduction}
Since the rediscovery of Bachelier's theory of the pricing of derivatives \cite{Bachelier} by Black, Scholes \cite{Black} and, independently, Merton \cite{Merton}, a lot of effort has been put into the research for generalizations of this pricing-formula. The Black-Scholes-Merton formula is based on the assumption that markets behave according to Gaussian diffusion. An alternative heuristic without Gaussian assumptions was worked out by Thorpe and Kassouf \cite{Thorpe} already in 1967, based on observations. Interesting enhancements of the Black-Scholes-Merton approach, either for option-pricing or interest-rate structure, have been published ever since, from which we only shall quote Heston \cite{Heston} and, more recently, Dragulescu \cite{Dragulescu}, who has calculated closed form solutions for a possible extension of Heston's model by the path-integral. Heston's model is interesting, since it shows that non-Gaussian behaviour can be achieved when the pricing-process is coupled to a volatility-process according to the Cox-Ingersoll-Ross model \cite{Cox}. Furthermore, it is known that Heston's model works far better than the Black-Scholes-Merton approach, see e.g. Tompkins \cite{Tompkins}. Note that the distribution-function of the Cox-Ingersoll-Ross model is Skellam's distribution \footnote{In more basic introductions to financial mathematics, e.g. \cite{Brigo}, the distribution-function of the Cox-Ingersoll-Ross model is often presented as a series of $\Gamma$- and $\chi^{2}$-distributions.}.

Investigations by Mandelbrot \cite{Mandelbrot} have shown, already in 1963, that markets do not behave like Gaussians, but show the fractal properties of L\'evy-distributions, L\'evians in all what follows. Mandelbrot has used exponentially truncated L\'evians for his examinations. The work of Cont. et. al. \cite{Cont} has shown that the behaviour of markets can be described best by a L\'evy-exponent $\alpha=1.7$\,, which is a small shift away from Gaussian behaviour, where $\alpha=2$\,. Pure L\'evians, however, have the disadvantage that they decay like power-laws, which does not allow for an easy generalization of the Black-Scholes-Merton formula as long as log-returns are considered. Exponentially truncated L\'evians, however, cannot be handled easily because, in order to exist mathematically, the range of the parameters has to be chosen carefully, see e.g. \cite{Paul}. 

Aguilar et. al. \cite{Aguilar} recently have worked out an analytic L\'evian pricing-formula that is based on the properties of the totally anti-symmetric L\'evian. The formula is based on highly sophisticated mathematics and, despite analytically available, does not allow for an easy implementation and application. 

Also recently, Kleinert et. al. \cite{Kleinert1}, \cite{Kleinert2} have worked out pricing formulas that are based on the path-integral and on double fractional diffusion, that is, that both time and space both behave fractionally. The generalization by Kleinert et. al. is achieved by using the integral-definitions of fractional derivatives, by which even skew behaviour can be implemented easily. Readers who seek more insight into the fractional world are referred to the exhaustive report by Metzler et. al. \cite{Metzler}.

A different ansatz has been proposed by Borland et. al. \cite{Borland1}, \cite{Borland2}. Borland et. al. use the fractal properties of Student's t-distribution which is also known as Tsallis's distribution. Like symmetric L\'evians, Tsallis's distribution has the advantage that it is steered by one clearly defined parameter $q$ only, whose numerical value for market-analysis is known to be $q=1.5$\,.

Our ansatz here pursues a different goal. We do not start with a differential-equation, but with the L\'evy-Khintchine theorem instead. This gives us the possibility to stay as close to the L\'evian structure as possible. We use the properties of the L\'evy-Khintchine theorem, see e.g. the textbook by Feller \cite{Feller}, a Markov-inequality and L\'evian noise to derive a fractional distribution-function that follows from an extremal-condition for the cumulant-function. The result is a fractional Gaussian that depends on the L\'evy-exponent $\alpha$ and has thus the advantage that only one clearly defined parameter enters. A generalized Gaussian has been worked out by Nadarajah \cite{Nadarajah}, too. However, our result is completely different in it's formal structure and in it's close relationship to the L\'evian.

Armed with our result we calculate option-pricing curves with the Black-Scholes-Merton formula. We find that a fair price of an option in a fractional market is to be considered higher than in an ideal Gaussian market. This, however, is to be expected, since fatter tails of the distribution-function give rise for larger movements of the market, such that a trader should pay more for this opportunity. However, once bought, the price-evolution of the option, if the underlying moves into the right direction, should also be higher than in an ideal Gaussian market. 

The difference between the Gaussian price and the fractional price can be understood by the concept of the implied volatility. We carry out a theoretical analysis of how the implied volatility would behave if market-prices would approximately behave like fractional Gaussian diffusion.

Because of it's generality, the fractional Gaussian may certainly be applied to other problems than option-pricing, too.

\section{Derivation of an extremal fractional Gaussian}
In this section we derive a distribution-function that has the same properties as a L\'evy-distribution, but decays exponentially. The way to achieve this goal is similar to the derivation of a probability bound. The resulting probability bound, with a proper normalization, but can of course be interpreted as a distribution-function. This distribution-function is extremal, because it is a consequence of a variational procedure.

At first we shall discuss the inverse problem of the L\'evy-Khintchine theorem, see e.g. \cite{Feller}. This will be the key-ingredient for the derivation of our extremal probability-distribution.
We suppose that we know the distribution-function $P(x)$ for some reason. Then we also know the cumulant-function $\Psi(k)$. The inverse problem is then given by the calculation of the noise-characteristic $\Delta(x)$ out of $\Psi(k)$. This connection is established by the L\'evy-Khintchine theorem. As a first step we write down the L\'evy-Khintchine theorem
\begin{equation}
\Psi(k)\,=\,\int_{-\infty}^{\infty}\,dx\,\left(\frac{e^{i\,k\,x}}{x^{2}}\,-\,\frac{1}{x^{2}}\,-\,i\,k\,\frac{\sin[x]}{x^{2}}\right)\,\Delta(x)\quad,
\label{1}\end{equation}
and observe, that the following equation holds
\begin{equation}
-\,\frac{\partial^{2}}{\partial\,k^{2}}\,\Psi(k)\,=\,\Delta(k)\,\leftrightarrow\,x^{2}\Psi(x)\,=\,\Delta(x)\quad.
\label{2}\end{equation}
The cumulant-function $\Psi(k)$ is thus governed by a Poisson-type equation in momentum space. In coordinate-space the relation is algebraic. The noise-characteristic in the L\'evian case is given by
\begin{equation}
\Delta(x)\,=\,\sigma_{\alpha}\,|x|^{2-\alpha}\quad.
\label{3}\end{equation}

For the next step towards an extremal distribution we need a Markov-inequality. The Markov-inequality makes a statement about the probability of a random-variable $X$ being larger or smaller than a certain value $x$.
The Markov inequality for the case of interest is given by
\begin{equation}
P(X\geq x)\,=\,P\left(e^{s\,X}\geq e^{s\,x}\right)\leq e^{-s\,x}\left<e^{s\,X}\right>\quad.
\label{4}\end{equation}
The exponentiation in fact corresponds to a Laplace-transform that establishes a connection to the characteristic function.

The characteristic function of a random process that is at least $N$-times divisible is given by
\begin{equation}
\left<e^{s\,X}\right>\,=\,\prod_{n=1}^{N}\left<e^{s\,X_{n}}\right>\,=\,\prod_{n=1}^{N}\sum_{\nu=0}^{\infty}\frac{s^{\nu}}{\nu!}\left<X_{n}^{\nu}\right>_{c}\quad,
\label{5}\end{equation}
where we have used the well-known connection to the cumulant-expansion. As we know from our treatment of the inverse problem above, Eq. (\ref{2}), the cumulant-function leaves the zeroth and the first moment free, such that the general form of the cumulant-function is given by
\begin{equation}
\prod_{n=1}^{N}\sum_{\nu=0}^{\infty}\frac{s^{\nu}}{\nu!}\left<X_{n}^{\nu}\right>_{c}
\,=\,\prod_{n=1}^{N}\left(1+s\left<X_{n}\right>_{c}+\frac{s^{2}}{2}\left<X_{n}^{2}\Psi(sX_{n})\right>\right)\quad.
\label{6}\end{equation}
The way to find a probability-bound for L\'evy distributions goes as follows. We have to start with the characteristic function
\begin{equation}
\left<e^{s\,X_{n}}\right>\,=\,1\,+\,s\left<X_{n}\right>\,+\,\frac{s^{2}}{2}\left<X_{n}^{2}\Psi(s\,X_{n})\right>\quad.
\label{7}\end{equation}
When we make direct use of our above result, Eq. (\ref{2}), we find
\begin{equation}
\left<e^{s\,X_{n}}\right>\,=\,1\,+\,s\left<X_{n}\right>\,+\,\frac{s^{2}}{2}\left<\Delta(s\,X_{n})\right>\quad.
\label{8}\end{equation}
The structure of a bound is introduced by re-exponentiating
\begin{equation}
\left<e^{s\,X_{n}}\right>\,\leq\,\exp\left[s\left<X_{n}\right>\,+\,\frac{s^{2}}{2}\,\left<\Delta(s\,X_{n})\right>\right]\quad,
\label{9}\end{equation}
and thus
\begin{equation}
\left<e^{s\,X}\right>\,\leq\,\exp\left[s\left<X\right>\,+\,\frac{s^{2}}{2}\,\sum_{n=1}^{N}\left<\Delta(s\,X_{n})\right>\right]\quad.
\label{10}\end{equation}
We now may approximate the last term in the exponent by
\begin{equation}
\sum_{n=1}^{N}\left<\Delta(sX_{n})\right>\,\approx\,\sigma_{\alpha}\,s^{2-\alpha}\quad.
\label{11}\end{equation}
Our last equation corresponds to the variance of the random-process and describes also all higher moments in a closed form. The exponent $\alpha$ still guarantees a fat but exponentially decaying tail.
\begin{figure}[t]\centering\vspace{0cm}
\rotatebox{0.0}{\scalebox{1.1}{\includegraphics{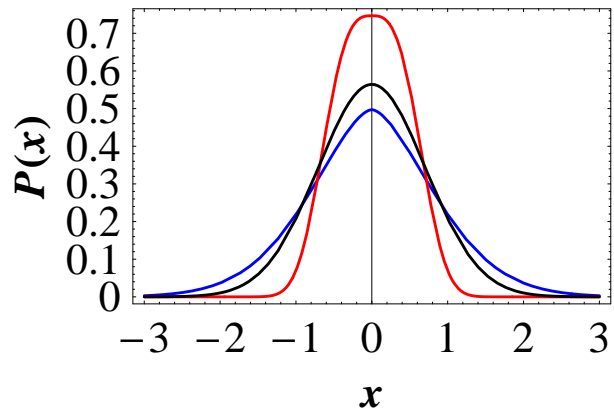}}}
\caption{\footnotesize{Illustraton of the dependency of Eq. (\ref{17}) on the exponent $\alpha$. The distributions are normalized.
The black graph ($\alpha=2$) is a Gaussian. The red graph ($\alpha=2.5$) illustrates the sub-Gaussian regime, while the blue graph ($\alpha=1.5$) illustrates the super-Gaussian regime. We have chosen $\sigma_{\alpha}=0.5$\,.}}
\label{generalizedgaussian1}\end{figure}
By putting all this together, we arrive at
\begin{equation}
\left<e^{s\,X}\right>\,\leq\,\exp\left[s\left<X\right>\,+\,\frac{\sigma_{\alpha}}{2}\,s^{4-\alpha}\right]\quad.
\label{12}\end{equation}
When we now use Markov's inequatlity Eq. (\ref{4}) and extremize the exponent of the probability bound in order to find the best possible value for $s$, we obtain
\begin{equation} 
s\,=\,\left(2\,|x|/(\sigma_{\alpha}\,(4-\alpha)\right)^{1/(3-\alpha)}\quad,
\label{13}\end{equation}
where we have set $\left<X\right>=0$. By interpreting the probability-bound as an extremal distribution-function, and by re-inserting $s$, we obtain
\begin{equation}
P(x;\, \alpha)
\,=\,N(\sigma_{\alpha},\,\alpha)\,\exp\left[-\,|x|\left(\frac{2\,|x|}{\sigma_{\alpha}\,(4-\alpha)}\right)^{\frac{1}{3-\alpha}}\right]\,
\exp\left[\left(\frac{2\,|x|}{\sigma_{\alpha}\,(4-\alpha)}\right)^{\frac{4-\alpha}{3-\alpha}}\frac{\sigma_{\alpha}}{2}\right]\quad.
\label{14}\end{equation} 
The best interpretation of our result is that of a fractional Gaussian. It is possible to calculate the normalization analytically,
\begin{equation}
N(\sigma_{\alpha},\,\alpha)\,=\,\frac{1}{2}
\left(\frac{3-\alpha}{4-\alpha}\left(\frac{2}{\sigma_{\alpha}\,(4-\alpha)}\right)^{\frac{1}{3-\alpha}}\right)^{\frac{3-\alpha}{4-\alpha}}\,\Gamma\left[\frac{2\,\alpha-7}{\alpha-4}\right]^{-1}\quad.
\label{15}\end{equation}

As it is elucidated by Fig.(\ref{generalizedgaussian1}), the extremal probability-distribution Eq. (\ref{14}) includes all possible transport-regimes in a physical sense. The Gaussian regime is properly reproduced for $\alpha = 2$\,, as it must. The super-Gaussian regime is given for $\alpha < 2$\,, as it should in order to be meaningful.
Additionally, we find that the sub-Gaussian regime is covered by the distribution-function Eq. (\ref{14}), too. The sub-Gaussian regime describes a transport that is extremely localized and thus approximately undisturbed by fluctuations. This case may also be interpreted as Ohmic transport. 
Thus, we have succeeded in deriving a distribution-function that depends on the L\'evy-exponent $\alpha$, that decays exponentially and that can be understood as a generalized fractional Gaussian.
To close this section, we shall note that the fractional Gaussian for
\begin{equation}
\lim_{\alpha\rightarrow3}P(x,\,\alpha)\,=\,\vartheta[1-|x|]\quad,
\label{16}\end{equation}
converges towards the step-function. As such, our result Eq. (\ref{14}) can also be understood as a distribution-function in a general sense.

Compared to an exponentially truncated L\'evy-flight our approach has the advantage that it works for any numerical values of the L\'evy-parameter. Conversely, exponentially truncated L\'evy-flights exist only for some specific numerical values of their parametric co-dimension, see e.g. \cite{Paul}.

\section{Option-pricing}
The Black-Scholes-Merton formula for an European Call-option can easily be generalized just by inserting the fractional Gaussian Eq. (\ref{14}) into the pricing-formula, where the function $G$ denotes the Greens-function of the distribution-function $P$\,,
\begin{equation}
\mathcal{C}(x,\,T-t;\,\alpha)\,=\,\int_{0}^{\infty}\,dy\,\left(e^{y}\,-\,1\right)\,G(x\,-\,y,\,\sigma_{\alpha}(T-t);\,\alpha)\quad.
\label{17}\end{equation}
The time-dependency enters as a product with the variance $\sigma_{\alpha}$\,. The time $T$ is the time of maturity. We have not succeeded in deriving a differential-equation for the fractional Gaussian, but it is safe to assume that the time-dependency has the same structure as in the Gaussian case. This, because we assume that the time-dependency is not fractional but behaves like ordinary diffusion. This assumption is supported by the fact that the fractional Gaussian Eq. (\ref{14}) is a direct consequence of the L\'evian structure of the system.
\begin{figure}[t]\centering\vspace{0cm}
\rotatebox{0.0}{\scalebox{0.82}{\includegraphics{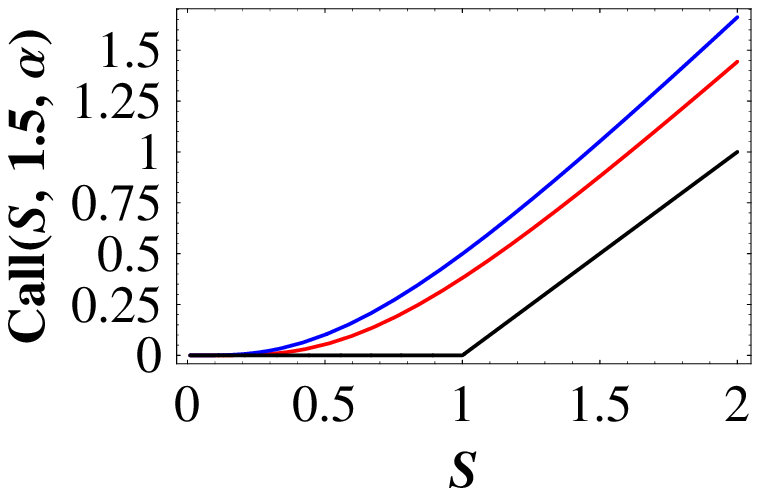}}}
\rotatebox{0.0}{\scalebox{0.84}{\includegraphics{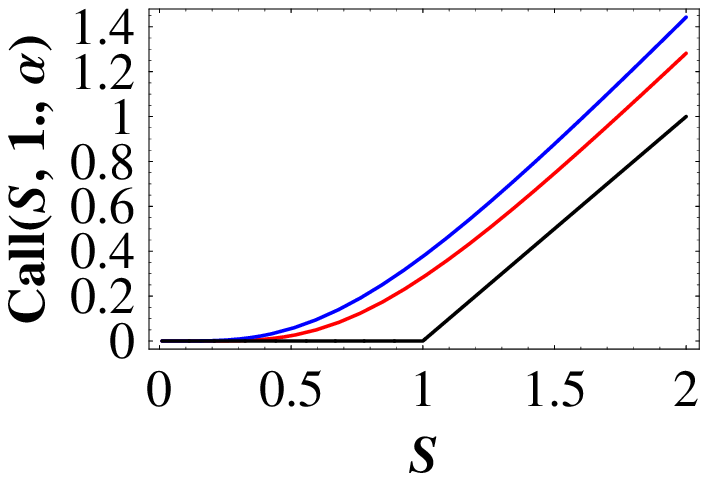}}}
\rotatebox{0.0}{\scalebox{0.82}{\includegraphics{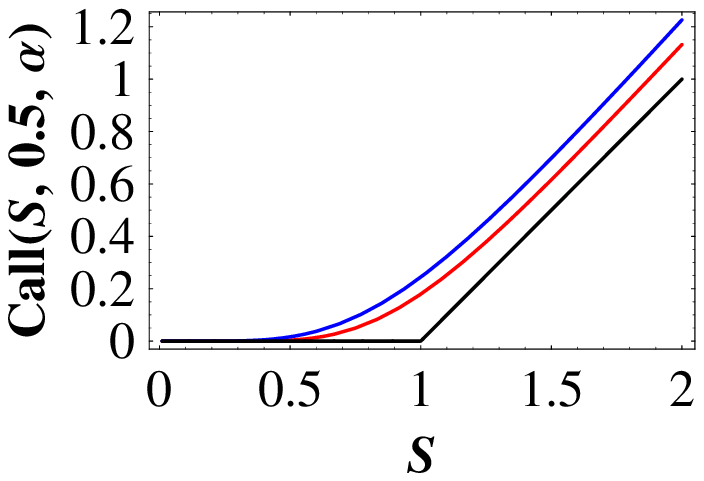}}}
\rotatebox{0.0}{\scalebox{0.82}{\includegraphics{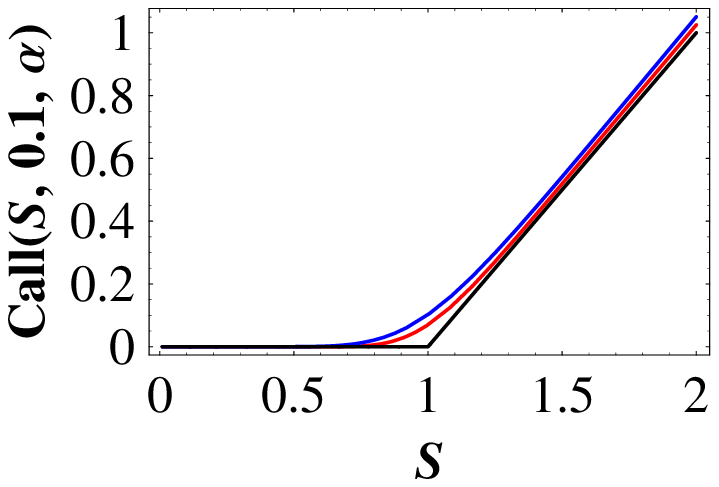}}}
\caption{\footnotesize{Comparison between the Black-Scholes-Merton price (red curve) and the price by the fractional Gaussian (blue curve) for four time-intervals to maturity, $\Delta t=\{1.5,\,1,\,05,\,0.1\}$\,. We have chosen $\alpha=1.7$ and $\sigma_{\alpha}=0.25$\,. The strike-price $K$ is chosen to be unity. The black line illustrates the maturity-curve.}}
\label{price}\end{figure}
In Fig. (\ref{price}) we have illustrated the prices as a function of $x=\ln[S]$. The blue curves illustrate the fractional price, while the red curves illustrate the Black-Scholes-Merton price. The choice $\alpha=1.7$ is due to the results of Cont et. al. \cite{Cont}. For the strike-price $K$ we have chosen unity. It can be deduced easily, that the fractional price is considerably larger than the Black-Scholes-Merton price. This is as it should, because a fatter distribution-function should generate higher prices. For the buyer this means that he should pay more, because a fatter tail enables the possibility of higher gains. For the holder, however, if the underlying asset evolves into the right direction, there should also be the possibility for a higher gains. For the issuer, a higher price for the option is of course an initial gain compared to the Black-Scholes-Merton price, however, there is also the risk of higher buy-back prices. Consequently, a trader can expect better chances than in a Gaussian market.

The difference between the fractional price and the Gaussian price in a market-situation is usually explained by the implied volatility instead of the L\'evy-exponent. The implied volatility then is given by the numerical value of the variance that has to be inserted into the Black-Scholes-Merton formula in order to approximately match the fractional price. A purely theoretical analysis of the behaviour of the implied volatility in our present approach can be carried out by the simulation of the fluctuations of prices by the fractional Gaussian $P(x;\, \alpha)$\,, and to calculate the volatility that should enter the Black-Scholes-Merton formula. This analysis can be carried out by the theorem on implicit functions. When we require that the Black-Scholes-Merton price equals the fractional price we must set
\begin{equation}
\mathcal{C}(S,\,\sigma_{\mathrm{imp}}(S,\,t),\,t,\,2)\,-\,\mathcal{C}(S,\,\sigma_{\alpha},\,t,\,\alpha)\,=\,0\quad,
\label{18}\end{equation}
where $\sigma_{\mathrm{imp}}(S,\,t)$ is the implied volatility as a function of the time $t$ and the price $S$. First order corrections to the volatility $\sigma_{\alpha}$ can now be calculated by solving Eq. (\ref{18}) for $d\sigma_{\mathrm{imp}}(S,\,t)$. By the implicit function theorem we obtain
\begin{equation}
\frac{d\sigma_{\mathrm{imp}}(S,\,t)}{d S}\,=\,\left(\frac{\partial\,\mathcal{C}(S,\,\sigma_{\alpha},\,t,\,2)}{\partial S}\,-\,\frac{\partial\,\mathcal{C}(S,\,\sigma_{\alpha},\,t,\,\alpha)}{\partial S}\right)\,\left(\frac{\partial\,\mathcal{C}(S,\,\sigma_{\alpha},\,t,\,2)}{\partial\sigma_{\alpha}}\right)^{-1}\quad.
\label{19}\end{equation}
The implied volatility to the first order can then be approximated by
\begin{equation}
\sigma_{\mathrm{imp}}(S,\,t)\,\approx\,\sigma_{\alpha}\,+\,d\sigma_{\mathrm{imp}}(S,\,t)\quad.
\label{20}\end{equation}
\begin{figure}[t]\centering\vspace{0cm}
\rotatebox{0.0}{\scalebox{1.1}{\includegraphics{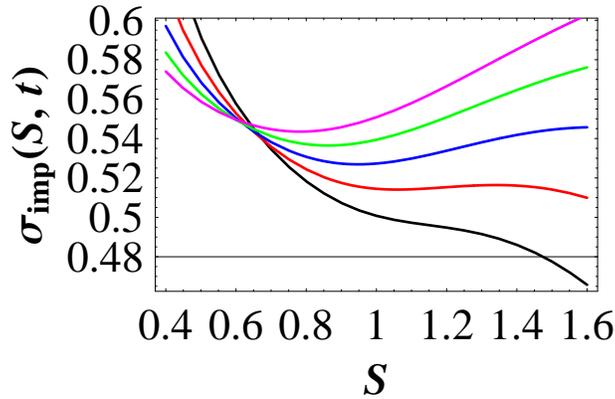}}}
\caption{\footnotesize{Implied volatilities for $\alpha=1.7$ and $\sigma_{\alpha}=0.4$\,. The curves are calculated for different time-units to maturity. Black ($\Delta t=0.5$)\,, red ($\Delta t=0.6$)\,, blue ($\Delta t=0.7$)\,, green ($\Delta t=0.8$) and purple ($\Delta t= 0.9$)\,. Smile and skew are clearly visible.}}
\label{volatility}\end{figure}
From Fig. (\ref{volatility}) it can be deduced that the implied volatility shows smile and skew behaviour. The purple and the green curve show the typical smile, however, the minimum does not lie at-the-money $S=1$, but somewhat below out-of-money. This is realistic since an option that has moved to shortly below at-the-money may certainly have the expectation to move in-the-money, such that it might pay off to wait in this situation. Waiting but means a slow-down in trading, which explains the fall-off of the volatility. Far in-the-money the fluctuations grow since trade becomes more intense again because traders might sell in order to cash in, or buy more because they have positive expectations. As maturity closes in, the fluctuations out-of-money grow, while the at-the-money point transforms into a saddle-point. High fluctuations out-of-money are realistic, since even close to maturity an out-of-money option still can mature in-the-money. The fall-off of the fluctuations in the in-the-money regime are understandable by that fact that all traders would naturally hold their options and wait for the option to mature. This would cause a slow-down of the market. We call the behaviour that we have yet described a roll-over. For different volatilities $\sigma_{\alpha}$ this effect takes place earlier or later. For smaller $\sigma_{\alpha}$ the roll-over takes place earlier, while for higher $\sigma_{\alpha}$ it occurs later. Later here means closer to maturity. This also seems to be realistic since in a market that shows high fluctuations the situation remains unclear even close to maturity, while with low fluctuations the situations seems to be clear even far from maturity. However, the behaviour of the market can change at any time, because locally the volatility $\sigma_{\alpha}$ and also the L\'evy-exponent $\alpha$ can be time-dependent. This then induces a cross-over from one volatility-curve to another. A cross-over is a clear additional risk, which certainly will drastically change the opinions of the traders and thus lead to different behaviour. Such effects are certainly self-organizing, since changes in $\sigma_{\alpha}$ and $\alpha$ change the implied volatility, which then changes the opinions of the traders, which then induces a back-reaction on $\sigma_{\alpha}$ and $\alpha$ and so on. Thus, real-market behaviour can very easily acquire a complexity that can be very demanding for every model.

Again we remind the reader that our simulation of the implied volatility here is purely theoretical. However, similar behaviour in smile and skew has also been found in the analysis by Borland et. al. \cite{Borland1, Borland2}, on the basis of a completely different approach.

\section{Conclusion}
By using the properties of the L\'evy-Khintchine theorem and a Markov-inequality we have derived an extremal fractional Gaussian that depends on the L\'evy-exponent $\alpha$. The fractional Gaussian mimics the behaviour of a L\'evian insofar, as that for $\alpha<2$ the shape becomes fatter and thus shows super-Gaussian behaviour. The draw-backs of the L\'evian, that is the power-law decay that prohibits convergence of the Black-Scholes-Merton formula but are eliminated, since the fractional Gaussian decays exponentially. This enables an easy implementation into the Black-Scholes-Merton formula. Other ans\"atze that have been mentioned in the introduction need more sophisticated mathematics in order to achieve convergence in this point.

Our approach shows that the price of an option in a fractional market should be higher than in a Gaussian market, because a fatter distribution-function enables higher gains for the buyer. This should be reflected in the price. The difference of the prices between a fractional and a Gaussian market may be explained by the implied volatility. 

We have carried out a theoretical analysis of how the implied volatility would behave if market-prices could approximately be described by the fractional Gaussian. In a real-world market environment, however, the behaviour of the implied volatility may certainly behave somewhat differently. This is, because in real markets it is likely that the underlying, or maybe hidden volatility $\sigma_{\alpha}$ is not a constant over time, such that cross-overs between different curves of the implied volatility can take place. We shall also mention that locally the L\'evy-exponent $\alpha$ might show a weak time-dependency, which would also change the behaviour of the implied volatility.

Because of it's generality the fractional Gaussian may certainly be applied to other problems of statistical analysis, too.


\end{document}